\documentclass[
  aps,
  pra,
  reprint,
  amsmath,amssymb,
  superscriptaddress,
  floatfix,
  longbibliography
]{revtex4-2}

\usepackage{graphicx}
\usepackage{bm}
\usepackage{amsmath}
\usepackage{mathtools}
\usepackage[T1]{fontenc}
\usepackage{xcolor}
\usepackage{siunitx}
\usepackage[colorlinks=true]{hyperref}
\definecolor{linknavy}{HTML}{1A3A6B}
\hypersetup{
  citecolor=linknavy,linkcolor=linknavy,urlcolor=linknavy,
  pdftitle={Schur-Horn bound on field-free molecular orientation at finite temperature},
  pdfauthor={},
  pdfkeywords={molecular orientation; rotational control; Schur-Horn bound; thermal ensemble}
}

\newcommand{\cost}{\cos\theta}
\newcommand{\avg}[1]{\langle #1 \rangle}
\newcommand{\tavg}[1]{\langle\!\langle #1 \rangle\!\rangle}
\newcommand{\Jmax}{J_{\mathrm{max}}}
\newcommand{\Trot}{T_{\mathrm{rot}}}
\newcommand{\kB}{k_{B}}
\newcommand{\Dexp}{D_{e}}
\providecommand{\ket}[1]{\lvert #1 \rangle}

\begin{document}

\title{Schur--Horn bound on field-free molecular orientation at finite temperature}

%
\author{Tanveer Ahmad}
\email{tanveer.quantum@gmail.com}
\affiliation{Hunan Key Laboratory of Super-Microstructure and Ultrafast Process, 
School of Physics, Central South University, Changsha 410083, China}

\date{\today}

\begin{abstract}
The maximum field-free orientation attainable from a \emph{thermal} molecular
ensemble within a finite rotational subspace has not been characterised
analytically. Here we derive a Schur--Horn-type upper bound on the field-free
orientation $\avg{\cos\theta}$ achievable by \emph{any} $M$-conserving unitary
control acting on a Boltzmann ensemble truncated to rotational levels $J\le\Jmax$.
The bound is the sum, over magnetic-quantum-number sectors, of the sorted Boltzmann
weights paired with the sorted spectrum of $\cos\theta$; it is purely kinematic, set
by temperature, the rotational constant, and $\Jmax$ alone, and interpolates exactly
between the zero-temperature subspace eigenvalue and a finite-temperature ceiling
fixed by the rotational partition function. Benchmarked on LiH against the analytical
$N$-subpulse resonant protocol of Hong \textit{et al.}
[Phys.\ Rev.\ Research \textbf{7}, L012049 (2025)], it reveals three regimes: the
protocol saturates the bound to within about $1.6\%$ for $T\le\SI{5}{\kelvin}$, loses
roughly $10\%$ of its zero-temperature orientation at
$T=B/\kB\approx\SI{10.8}{\kelvin}$, and leaves a $10$--$40\%$ gap above
$\SI{10}{\kelvin}$. Optimising the subpulse areas and carrier phases within the fixed
layout closes only $\approx7\%$ of that gap, the optimal phase offsets vanishing,
which localises the dominant loss in the rigidity of the analytic layout rather than
in the choice of areas or phases. The bound is a control-independent target for
coherent-control design, applies unchanged to any linear polar molecule in a
$^1\Sigma^+$ state, and is mapped across the $(\Jmax,T)$ plane.
\end{abstract}

\maketitle

\section{Introduction}
\label{sec:intro}

Field-free molecular orientation, the preferential alignment of the molecular
dipole along a laboratory axis after the controlling field has been switched
off, is a prerequisite for a broad class of experiments, from stereochemistry and
high-harmonic generation in polar molecules to attosecond electron-recollision
imaging and the precision measurement of fundamental
symmetries~\cite{StapelfeldtSeideman2003,KochLemeshkoSugny2019}. Precise control of
molecular rotation also underpins proposals to encode quantum information in the
rotational states of polar molecules, as qubits and higher-dimensional
qudits~\cite{DeMille2002,Albert2020,Sawant2020}. A wide range of
control strategies has been developed, including half-cycle
pulses~\cite{MachholmHenriksen2001}, two-colour femtosecond
fields~\cite{KanaiSakai2001,WangHenriksen2020}, intense single-cycle terahertz
pulses~\cite{Fleischer2011,Egodapitiya2014,Xu2020}, optimally shaped or kicked
fields~\cite{Sugny2004,Daems2005}, and resonant microwave or
terahertz pulse trains~\cite{Salomon2005,Karras2015,ShuHenriksen2013}. These methods
have realised orientation from the two-state limit~\cite{Trippel2015} to all-optical
three-dimensional control of asymmetric tops~\cite{Lin2018} and cavity-mediated
control of a single molecular polariton~\cite{Fan2023}. Resonant
schemes that climb the rotational ladder are especially attractive: they operate at
modest field strengths and at carrier frequencies that are fully under experimental
control, and they admit closed-form design. Indeed, for a molecule prepared in the
rotational ground state $\ket{0,0}$, an explicit analytical prescription now exists
that maximises the orientation within a chosen truncated rotational basis: the
$N$-subpulse protocol of Hong \textit{et al.}~\cite{Hong2025} fixes the target state
to the principal eigenvector of $\cos\theta$ in the $\ket{J,0}$ subspace and
realises it with a sequence of resonant subpulses of analytically prescribed areas,
phases, and timings.

What is missing is a statement of the \emph{fundamental limit}. Any real ensemble is
thermal, not a pure $\ket{0,0}$ ket, and the orientation that a thermal ensemble can
reach is constrained not by a particular pulse design but by the rotational
states available within the controllable subspace and by the spectrum of the
orientation operator itself. The relevance of this limit is not academic:
preparation techniques deliver polar molecules over an enormous range of rotational
temperatures, from $\sim\SI{100}{\micro\kelvin}$ in direct laser
cooling~\cite{Anderegg2017,Truppe2017} through a few kelvin in buffer-gas
cooling~\cite{PattersonDoyle2007}, Stark deceleration~\cite{Bethlem2000}, and
optoelectrical cooling~\cite{Prehn2016}, up to tens of kelvin in supersonic-jet
sources, so that for LiH (rotational constant $B\approx\SI{10.8}{\kelvin}$ in
temperature units) the dimensionless parameter $\kB T/B$ spans some six orders of
magnitude. Thermal averaging is moreover the single largest correction to ideal
orientation predictions in existing terahertz
experiments~\cite{Fleischer2011,Xu2020}. Finite-temperature orientation has itself
been analysed for specific control fields, both for thermal ensembles of
$^{1}\Sigma$ and $^{2}\Sigma$ molecules~\cite{Tehini2012} and for the maximal
orientation attainable with optimal-control terahertz pulses~\cite{Liao2013}; what
these protocol-specific studies leave open is a \emph{control-independent} ceiling
that every realisation must respect. Such a bound would therefore serve both as a
target for protocol design and as a diagnostic that separates intrinsic
thermodynamic limits from controllable inefficiencies.

\begin{figure*}[t]
  \centering
  \includegraphics[width=0.96\linewidth]{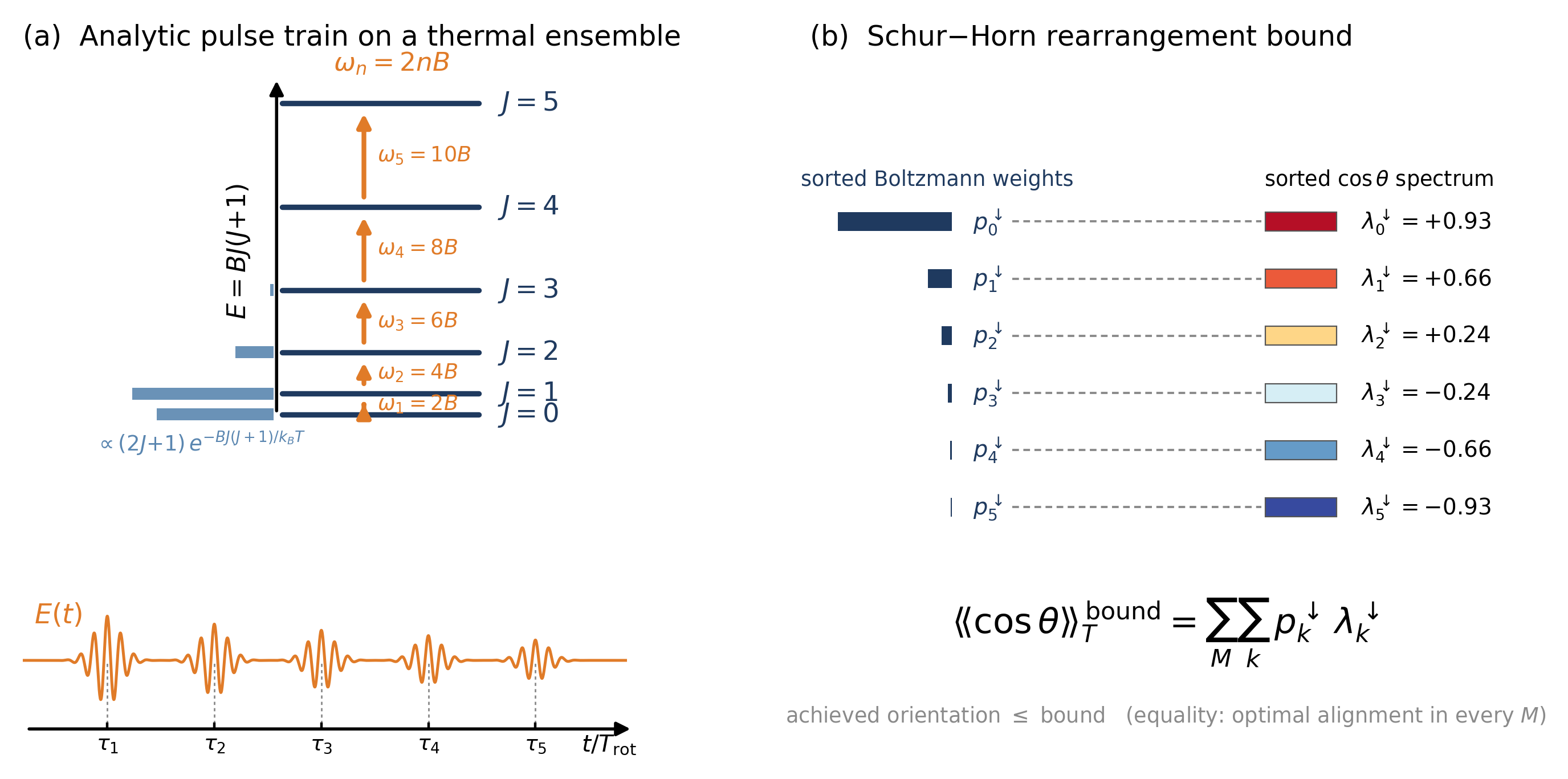}
  \caption{Schur--Horn bound on thermal orientation. (a) A linearly polarised
  $N$-subpulse field $E(t)$ resonantly climbs the rotational ladder
  ($\omega_n=2nB$), acting on a Boltzmann-populated ensemble (left bars). Because the
  field conserves $M$, the dynamics block-diagonalise into magnetic-quantum-number
  sectors. (b) Within each $M$-sector the maximal field-free orientation attainable
  by any unitary control is the inner product of the Boltzmann weights sorted in
  decreasing order, $\{p^{\downarrow}_k\}$, with the $\cos\theta$ eigenvalues sorted
  in decreasing order, $\{\lambda^{\downarrow}_k\}$ (Schur--Horn / rearrangement
  inequality). Summing over sectors gives the bound of Eq.~\eqref{eq:bound}. Any
  physical pulse train induces a single evolution and can only reach this ceiling if
  its unitary realises the optimal pairing in every sector simultaneously.}
  \label{fig:schematic}
\end{figure*}

In this work we provide such a bound. Our central result is an analytical
Schur--Horn-type upper bound on the field-free orientation achievable by any
$M$-conserving unitary control acting on a finite-temperature rotational ensemble
truncated to $J\le\Jmax$ [Eq.~\eqref{eq:bound}]. The construction is summarised in
Fig.~\ref{fig:schematic}: a linearly polarised field conserves the magnetic quantum
number $M$, so the thermal state block-diagonalises into $M$-sectors, and within
each sector the rearrangement (Schur--Horn) inequality caps the attainable
$\avg{\cos\theta}$ by the inner product of the sorted Boltzmann weights with the
sorted $\cos\theta$ eigenvalues. Summed over sectors, this yields a closed,
control-independent ceiling that interpolates exactly between the zero-temperature
subspace eigenvalue $\lambda$ and a finite-temperature limit fixed by the rotational
partition function. We then (i) validate a numerically exact $M$-sector propagator
against the analytic zero-temperature protocol; (ii) compute the thermal orientation
delivered by the analytic protocol on LiH across two decades in temperature; (iii)
benchmark it against the bound, identifying three temperature regimes; (iv) construct
an operability map, i.e.\ the achievable orientation across the $(\Jmax,T)$ plane; and (v) show, by direct optimisation,
that retuning the subpulse areas \emph{and} carrier phases within the fixed pulse-train
layout recovers only a small fraction of the thermal shortfall (the optimal phase
offsets vanishing), which localises the dominant loss in the
\emph{structure} of the analytic ansatz.

The paper is organised as follows. Section~\ref{sec:theory} sets up the rotational
control problem, the thermal ensemble in the $M$-conserving basis, and derives the
Schur--Horn bound. Section~\ref{sec:results} presents the numerical validation,
the thermal benchmarking against the bound, the operability map, the fixed-layout
optimisation, and the experimental implications and limitations.
Section~\ref{sec:conclusion} concludes.

\section{Theoretical framework}
\label{sec:theory}

\subsection{Rotational control problem}

The rotational Hamiltonian of a polar diatomic molecule in its $^1\Sigma^+$
electronic state, neglecting centrifugal distortion and addressed by a linearly
$\hat z$-polarised field $E(t)$, is
\begin{equation}
  H(t) \;=\; B\,\hat{\bm{J}}^2 \;-\; \mu_0\,E(t)\,\cost ,
  \label{eq:hamiltonian}
\end{equation}
where $B$ is the rotational constant, $\mu_0$ the permanent dipole, and $\theta$ the
polar angle of the internuclear axis relative to $\hat z$. In the $\ket{J,M}$ basis
the field-free Hamiltonian is diagonal with eigenvalues $B\,J(J+1)$, and the dipole
operator couples only $J'=J\pm1$ at fixed $M$, with the matrix elements
\begin{equation}
  \langle J+1,M|\cost|J,M\rangle
  \;=\; \frac{\sqrt{(J+1)^2-M^2}}{\sqrt{(2J+1)(2J+3)}} .
  \label{eq:dipole}
\end{equation}
The control objective is to maximise the field-free expectation value
$\avg{\cost}$ at a target time after the field has terminated; only its magnitude
$|\avg{\cost}|$ is physically distinguished, since a global $\pi$ shift of the
carrier phases reverses the sign of the orientation without changing its size, and
we report the positive value (dipole along $+\hat z$) throughout. For a pure $\ket{0,0}$
initial state confined to the $(\Jmax{+}1)$-dimensional $\ket{J,0}$ subspace, the
optimum is the principal eigenvalue $\lambda$ of the truncated operator
$[\cost]_{\mathrm{sub}}$, attained on its principal eigenvector
$\bm{c}^{\star}$~\cite{Hong2025}. The questions we address are how this optimum is
modified for a thermal ensemble, and how close a realisable control field comes to
it.

\subsection{Analytic \texorpdfstring{$N$}{N}-subpulse control protocol}
\label{sec:protocol}

As a concrete, state-of-the-art realisable control we adopt the analytic protocol of
Ref.~\cite{Hong2025}, which we summarise here because it is the design we benchmark
against the bound. The field is a train of $N=\Jmax$ non-overlapping resonant
subpulses of common Gaussian envelope,
\begin{equation}
  E(t) \;=\; \sum_{n=1}^{N} F_n^{(0)}\,
  e^{-(t-\tau_n)^2/(2T_p^2)}\,
  \cos\!\big[\omega_n (t-\tau_n)+\phi_n\big],
  \label{eq:field}
\end{equation}
with central times $\tau_n=(n-1)\,5T_p$, carrier frequencies $\omega_n=2nB$ resonant
with the $\ket{n-1,0}\to\ket{n,0}$ transition, and rms width $T_p=3\Trot$ (so that
each subpulse is spectrally narrow enough to address a single rotational transition:
the nearest neighbour, detuned by $2B$, lies $2BT_p=6\pi\approx19$ spectral widths
away), where
$\Trot=\pi/B$. The target amplitudes $c^{\star}_J$ that maximise
$\sum_{J,J'} c^{\star}_J\langle J|\cost|J'\rangle c^{\star}_{J'}$ satisfy
\begin{equation}
  [\cost]_{\mathrm{sub}}\,\bm{c}^{\star} \;=\; \lambda\,\bm{c}^{\star},
  \label{eq:eigenvalue}
\end{equation}
with $\lambda$ the largest eigenvalue of the tridiagonal $\cos\theta$ matrix (or,
equivalently, the largest positive root of its characteristic polynomial). The
subpulse areas follow from
\begin{equation}
  \theta_1=\arccos c^{\star}_0,\qquad
  \theta_n=\arccos\!\left[\frac{c^{\star}_{n-1}}
  {\prod_{k=1}^{n-1}\sin\theta_k}\right]\quad (n\ge 2),
  \label{eq:areas}
\end{equation}
together with the carrier phases $\phi_n\equiv\pi/2+\omega_n\tau_n\ (\mathrm{mod}\ 2\pi)$
and peak amplitudes
\begin{equation}
  F_n^{(0)} \;=\; \sqrt{\tfrac{2}{\pi}}\;\frac{\theta_n}{\mu_n T_p},
  \qquad
  \mu_n=\mu_0\,\frac{n}{\sqrt{(2n+1)(2n-1)}} .
  \label{eq:amplitude}
\end{equation}
Here the pulse area is defined within the rotating-wave approximation as
$\theta_n=\tfrac12\mu_n F_n^{(0)}\sqrt{2\pi}\,T_p$, i.e.\ the time integral of the
co-rotating Rabi frequency $\tfrac12\mu_n F_n(t)$ over the Gaussian envelope; the
prefactor $\sqrt{2/\pi}$ in Eq.~\eqref{eq:amplitude} follows from this convention
together with $T_p$ (since $\int e^{-t^2/2T_p^2}dt=\sqrt{2\pi}\,T_p$, and the factor
$\tfrac12$ is the rotating-wave projection of the linearly polarised
carrier).
After the train, the orientation revives at every integer multiple of $\Trot$; we
evaluate it at the first full revival, $t_f\approx 222.5\,\Trot$ for this layout. The
absolute time origin and phase convention do not affect the peak orientation.

\subsection{Thermal ensemble in the \texorpdfstring{$M$}{M}-conserving basis}

Because a linearly $\hat z$-polarised field conserves $M$, the initial Boltzmann
ensemble, diagonal in $(J,M)$, evolves sector by sector, and the thermal
expectation factorises:
\begin{equation}
  \tavg{\cost}_T(t)=\frac{1}{Z}
  \sum_{M=-\infty}^{\infty}\!\sum_{J\ge|M|}\!
  e^{-BJ(J+1)/\kB T}\!\avg{\cost}_{J,M}(t),
  \label{eq:thermal}
\end{equation}
where $Z=\sum_J (2J+1)\,e^{-BJ(J+1)/\kB T}$ is the rotational partition function and
$\avg{\cost}_{J,M}(t)$ is the expectation when the system begins in $\ket{J,M}$ and
is driven by a given field. For $M\neq0$ the dipole couplings of
Eq.~\eqref{eq:dipole} are systematically smaller than at $M=0$; the effective Rabi
frequency of a subpulse acting on $\ket{J,M}$ is reduced relative to the $M=0$ value
used to design the protocol by
\begin{equation}
  \eta(J,M)\;=\;\sqrt{1-\frac{M^2}{(J+1)^2}} .
  \label{eq:eta}
\end{equation}
Since $\pm|M|$ have identical $\cos\theta$ dynamics, Eq.~\eqref{eq:thermal} reduces to
a sum over $M\ge0$ with multiplicity $2$ for $M\neq0$ and $1$ for $M=0$; this is the
form used throughout.

\subsection{Schur--Horn upper bound}
\label{sec:bound}

We now bound the orientation attainable by \emph{any} control, not only the protocol
of Sec.~\ref{sec:protocol}. For a fixed truncation
$\mathcal{S}_{\Jmax,M}=\{\ket{J,M}:|M|\le J\le\Jmax\}$ of dimension
$d_M=\Jmax-|M|+1$, let $\rho_M(T)$ be the restriction of the thermal state to that
subspace and $C_M$ the restriction of $\cos\theta$. The maximum of
$\mathrm{tr}[\rho_M(T)\,U_M^{\dagger} C_M U_M]$ over all unitaries $U_M$ acting within
$\mathcal{S}_{\Jmax,M}$ is given by the Schur--Horn / von Neumann rearrangement
inequality~\cite{Horn1954,Bhatia1997,MarshallOlkin2011},
\begin{equation}
  \max_{U_M}\,\mathrm{tr}\!\big[\rho_M(T)\,U_M^{\dagger} C_M U_M\big]
  \;=\;\sum_{k=0}^{d_M-1} p^{(M)\downarrow}_k\,\lambda^{(M)\downarrow}_k,
  \label{eq:msector-bound}
\end{equation}
where $\{p^{(M)\downarrow}_k\}$ and $\{\lambda^{(M)\downarrow}_k\}$ are the eigenvalues
of $\rho_M(T)$ and $C_M$, each sorted in decreasing order. The maximum is saturated by
the unitary that maps the principal eigenvector of $\rho_M$ onto the principal
eigenvector of $C_M$, the second onto the second, and so on. Summing over $M$ yields a
bound on the achievable thermal orientation:
\begin{equation}
  \tavg{\cost}_T^{\,\mathrm{bound}}(\Jmax)\;=\;
  \sum_{|M|\le\Jmax}\;\sum_{k=0}^{d_M-1}
  p^{(M)\downarrow}_k\,\lambda^{(M)\downarrow}_k,
  \label{eq:bound}
\end{equation}
normalised through $Z$ so that $\sum_{M,k} p^{(M)}_k=1$. Equation~\eqref{eq:bound} is
the central analytical result of this paper [Fig.~\ref{fig:schematic}(b)]. Four
remarks fix its meaning.

\emph{(i) Tightness within $M$.} Within each $M$-sector, Eq.~\eqref{eq:msector-bound}
is exactly saturated by the rearranging unitary, so the bound is the best possible
$\avg{\cost}$ within $\mathcal{S}_{\Jmax,M}$ assuming an independent control unitary
per sector.

\emph{(ii) Looseness across $M$.} A physical pulse $E(t)$ induces a \emph{single}
evolution which, restricted to different sectors, generally produces different
unitaries $U_M$ (the couplings depend on $M$). Equation~\eqref{eq:bound} optimises
each sector independently and is therefore an upper bound on what any one pulse train
can achieve; its tightness measures how compatible the sectors' demands are.

\emph{(iii) Zero-temperature limit.} As $T\to0$, $\rho_M$ collapses to the projector
onto $\ket{0,0}$ in the $M=0$ sector and to zero elsewhere, so
Eq.~\eqref{eq:bound} reduces to $\lambda^{(M=0)}=\lambda$, recovering the pure-state
optimum.

\emph{(iv) Subspace dependence and kinematic character.} Equation~\eqref{eq:bound} is
bounded above by the largest eigenvalue of $\cos\theta$ in the untruncated basis,
which is unity, and interpolates monotonically between $\lambda$ at $T\to0$ and a
value strictly below unity at finite $T$ and finite $\Jmax$. Crucially, the bound
depends only on the Boltzmann weights and the $\cos\theta$ spectrum (both purely
kinematic) and not on the field or on the rotational \emph{energies}; it is therefore
unaffected by energy-level corrections such as centrifugal distortion
(Sec.~\ref{sec:results}\,F).

\section{Results and discussion}
\label{sec:results}

\subsection{Validation of the numerical method}
\label{sec:validation}

The thermal benchmark below relies on an $M$-sector propagator, which we first
validate against the analytic zero-temperature protocol of Sec.~\ref{sec:protocol}.
We adopt the LiH parameters $B=\SI{7.513}{\per\centi\meter}$ and $\mu_0=\SI{5.88}{D}$,
giving $\Trot\approx\SI{2.22}{\pico\second}$ and $B/\kB\approx\SI{10.81}{\kelvin}$.
The time-dependent Schr\"odinger equation is propagated in the $\ket{J,0}$ basis up to
$J=20$ (five levels above $\Jmax=15$) with a second-order Strang
split-operator scheme~\cite{FeitFleckSteiger1982},
\begin{equation}
  U(\Delta t)\;=\;
  e^{-iH_0\Delta t/2}\;
  e^{+i\mu_0 E(t_m)\cost\,\Delta t}\;
  e^{-iH_0\Delta t/2},
  \label{eq:propagator}
\end{equation}
with midpoint $t_m$ and $\Delta t\approx\SI{38}{a.u.}$
($\Delta t/\Trot\approx4\times10^{-4}$). The $+i$ sign follows from the interaction
term $-\mu_0 E\cost$. We implemented the propagator independently in Python and MATLAB
and confirmed agreement to single-precision tolerance.

Table~\ref{tab:validation} reports eight quantitative checks for
$\Jmax=1,\dots,15$. The eigenvalue from the characteristic polynomial matches direct
diagonalisation; the recursion~\eqref{eq:areas} reproduces the eigenvector to machine
precision; numerical integration recovers each designed area; the simulated peak
orientation reaches $\lambda$ to $1.2\times10^{-4}$, limited by the time step; and
population leakage out of the design subspace is $\sim10^{-9}$, confirming the basis
buffer is more than adequate. The propagated pulse sequences and the time traces of
$\avg{\cost}(t)$ and $\avg{\cos^2\theta}(t)$ reproduce the analytic design within these
tolerances. We adopt the phase convention $\phi_n=\pi/2+\omega_n\tau_n$, which places
the orientation maxima at integer $\Trot$; this is a choice of phase reference and does
not affect the peak values.

\begin{table}[t]
\caption{Validation of the $M$-sector propagator against the analytic zero-temperature
design, for $\Jmax=1,\dots,15$. Each row reports the maximum error over that range.
M.p.\ $=$ machine precision.}
\label{tab:validation}
\begin{ruledtabular}
\begin{tabular}{lr}
Test & Max.\ error \\
\colrule
Eq.~\eqref{eq:eigenvalue} eigenvalue: \texttt{poly/roots} vs.\ \texttt{eigh} & $5\times10^{-13}$ \\
Eigenvector $\|\bm{c}^{\star}\|$ \& $\mathrm{sign}(c^{\star}_0)$ & $<10^{-10}$, M.p. \\
Eq.~\eqref{eq:areas} vs.\ numerical $\bm{c}^{\star}$ & $7\times10^{-15}$ \\
Pulse-area $\int$ vs.\ designed $\theta_n$ & $5\times10^{-16}$ \\
Peak-orientation timing vs.\ integer $\Trot$ & $5\times10^{-4}\,\Trot$ \\
Leakage to $J>\Jmax$ at $t_f$ & $9\times10^{-10}$ \\
$\max\avg{\cos^2\theta}$ vs.\ $\bm{c}^{\star\top}\cos^2\theta\,\bm{c}^{\star}$ & $2\times10^{-4}$ \\
$\max\avg{\cost}$ vs.\ $\lambda$ & $1.2\times10^{-4}$ \\
\end{tabular}
\end{ruledtabular}
\end{table}

\subsection{Thermal response of the analytic protocol}
\label{sec:thermal-response}

We evaluate the thermal orientation of Eq.~\eqref{eq:thermal} on a temperature grid
$T\in\{0,0.1,0.5,1,2,5,10,20,50\}\,\si{\kelvin}$ for designs
$\Jmax\in\{1,3,5,7,10,13,15\}$, propagating all initial states $\ket{J_0,M}$ with
$|M|\le12$ and $J_0\le18$ in parallel. The omitted Boltzmann weight at the basis edge
is $e^{-18\cdot19\,B/\kB T}\approx8\times10^{-33}$ at the highest temperature, so the
truncation is harmless (Appendix~\ref{app:numerics}).

\begin{figure}[t]
  \centering
  \includegraphics[width=\linewidth]{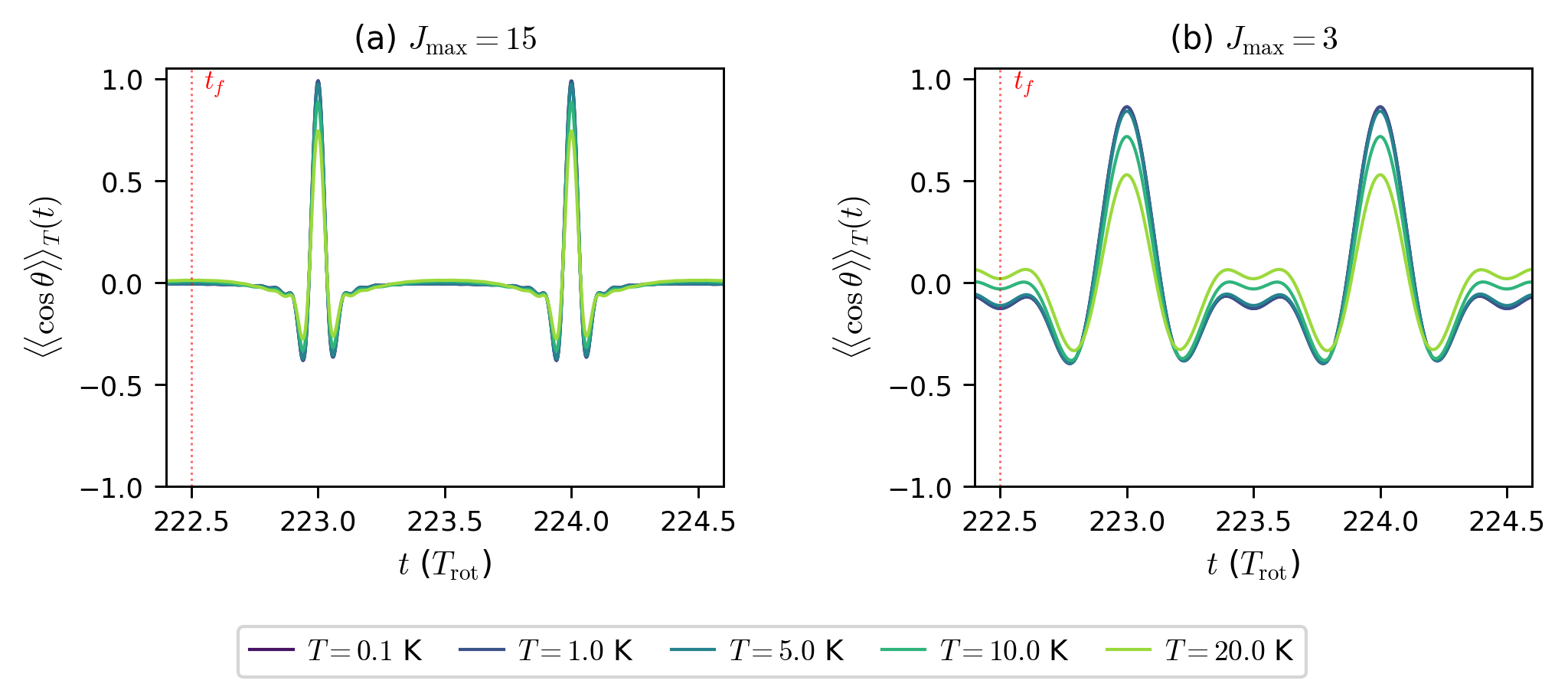}
  \caption{Field-free thermal orientation $\tavg{\cost}_T(t)$ near the reference time
  $t_f$ for several rotational temperatures: (a) $\Jmax=15$ and (b) $\Jmax=3$, both for
  the same protocol designed at $T=0$. The revivals at integer $\Trot$ are preserved at
  low $T$ and progressively dampen and broaden as $T$ rises. For $\Jmax=15$ the peak
  drops from $0.989$ at $T\to0$ to $0.74$ at $\SI{20}{\kelvin}$; for $\Jmax=3$ the
  relative loss is much larger at the same $T$ ($61\%$ of the zero-temperature value
  retained at $\SI{20}{\kelvin}$, $31\%$ at $\SI{50}{\kelvin}$) because the train cannot
  reorient population above $J=3$.}
  \label{fig:thermal-traces}
\end{figure}

Figure~\ref{fig:thermal-traces} shows the thermal orientation near $t_f$ for two
representative designs. Below $\sim\SI{1}{\kelvin}$ the traces are indistinguishable
from the zero-temperature result. By $\SI{5}{\kelvin}$ the revival amplitude is
visibly reduced; at $\SI{20}{\kelvin}$ the $\Jmax=15$ design still produces a clean
peak at each integer $\Trot$ but with amplitude reduced from its zero-temperature
value $0.989$ to $0.74$ (a $\sim25\%$ loss), while the $\Jmax=3$ design degrades far
faster, retaining only $61\%$ of its zero-temperature orientation at $\SI{20}{\kelvin}$
and $31\%$ at $\SI{50}{\kelvin}$. The reason is structural: the $\Jmax=3$ train can
reorient population only up to $J=3$, so as the temperature rises and higher-$J$ levels
are populated, the unaddressed weight, compounded by the $M$-dependent area mismatch in
the occupied $J\le3,\,|M|\ge1$ levels, progressively erodes the thermal average.

\begin{figure}[t]
  \centering
  \includegraphics[width=\linewidth]{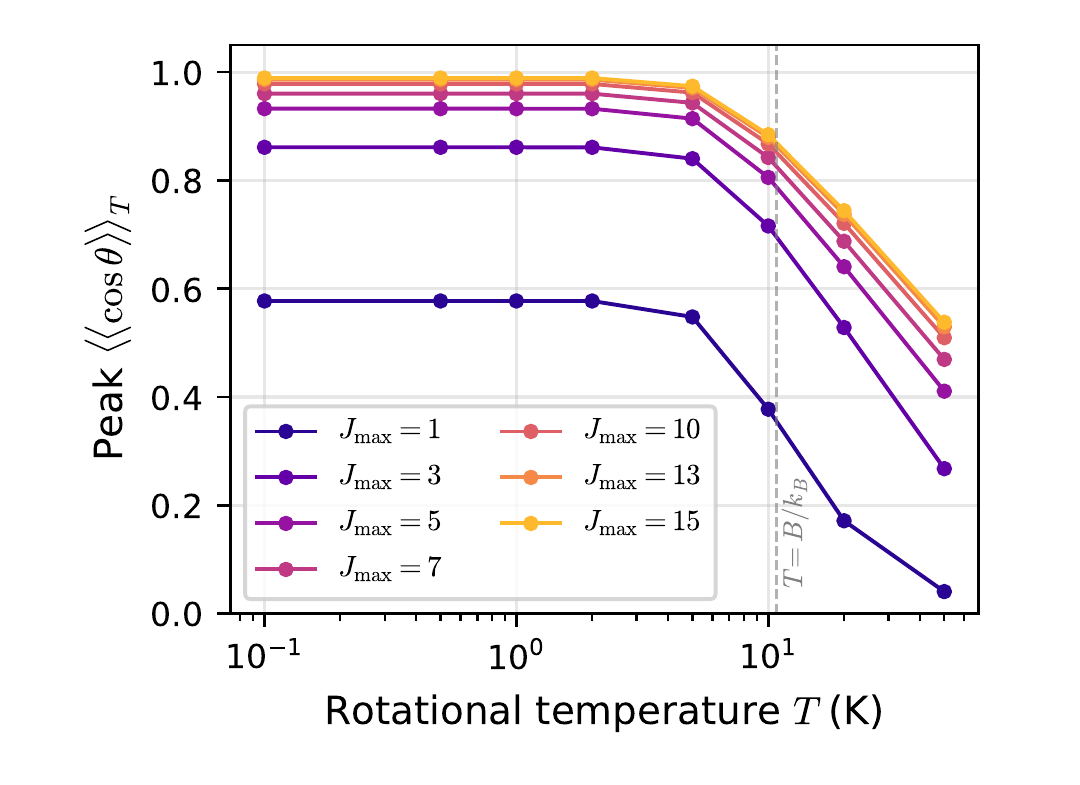}
  \caption{Peak thermal orientation $\max_t\tavg{\cost}_T(t)$ versus temperature for
  $\Jmax=1,3,5,7,10,13,15$. The dashed vertical line marks
  $T=B/\kB\approx\SI{10.8}{\kelvin}$. All curves are flat for $T\lesssim\SI{2}{\kelvin}$;
  the higher-$\Jmax$ protocols are more robust (at $\SI{10}{\kelvin}$, $\Jmax=15$ retains
  $0.88$ while $\Jmax=1$ has collapsed to $0.38$).}
  \label{fig:peak-vs-T}
\end{figure}

Figure~\ref{fig:peak-vs-T} maps the peak orientation against temperature for all seven
designs. Every curve stays at its zero-temperature value (the eigenvalue $\lambda$) up
to $T\approx\SI{2}{\kelvin}$, passes through an inflection near
$T=B/\kB\approx\SI{10.8}{\kelvin}$, and declines as the population spreads. The
higher-$\Jmax$ protocols are markedly more thermally robust, for a structural reason:
each subpulse $n$ resonantly drives $\ket{n-1,M}\to\ket{n,M}$ for every $|M|<n$, so a
$\Jmax=15$ train directly addresses all $\ket{J,M}$ with $J\le14$ and $|M|\le14$
($>99\%$ of the LiH population at any $T\le\SI{50}{\kelvin}$), whereas a $\Jmax=1$ train
addresses only $\ket{0,0}\to\ket{1,0}$.

\subsection{Benchmarking against the Schur--Horn bound}
\label{sec:benchmark}

\begin{figure*}[t]
  \centering
  \includegraphics[width=0.92\linewidth]{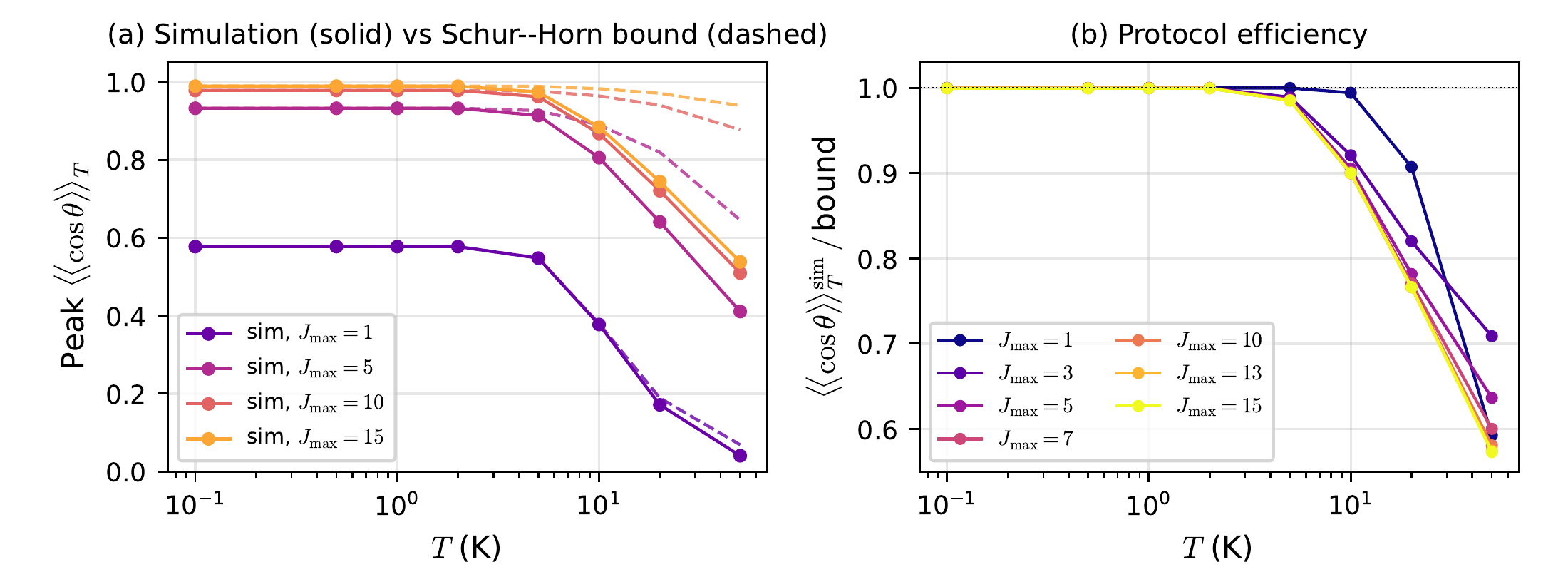}
  \caption{(a) Peak thermal orientation from the analytic protocol (solid, full
  symbols) and the Schur--Horn upper bound of Eq.~\eqref{eq:bound} (dashed) for
  $\Jmax=1,5,10,15$. Below $\sim\SI{2}{\kelvin}$ the protocol saturates the bound; above
  $\sim\SI{10}{\kelvin}$ a gap opens that grows with $T$ and $\Jmax$. (b) Ratio of
  simulated orientation to the bound for all seven designs: $\gtrsim0.984$ for
  $T\le\SI{5}{\kelvin}$, then a sharp drop, with the high-$\Jmax$ protocols retaining
  $\sim58\%$ at $\SI{50}{\kelvin}$. The gap quantifies the head-room available, in
  principle, to any improved control.}
  \label{fig:sim-vs-bound}
\end{figure*}

Figure~\ref{fig:sim-vs-bound} compares the simulated peak orientation with the bound of
Eq.~\eqref{eq:bound}, revealing three regimes.

\emph{Regime I ($T\lesssim\SI{2}{\kelvin}$, $\kB T\lesssim0.2\,B$).} The $J>0$
population is $\le0.03\%$ ($e^{-2B/\kB T}\le10^{-4}$); the thermal state is effectively
the pure $\ket{0,0}$ state, for which the protocol is optimal, and it saturates the
bound (ratio $\ge0.9999$).

\emph{Regime II ($\SI{2}{\kelvin}\lesssim T\lesssim\SI{5}{\kelvin}$).} The $J=1$
multiplet acquires up to $\sim10\%$ of the population, but largely in $M=0,\pm1$ levels
that remain well coupled. The simulation-to-bound ratio stays above $0.984$ and the
protocol is close to optimal.

\emph{Regime III ($T\gtrsim\SI{10}{\kelvin}$, $\kB T\gtrsim B$).} Substantial population
reaches $J\ge2,\,|M|\ge2$. The protocol drives each transition with the area
appropriate to $M=0$, but the $|M|\ge1$ couplings are smaller by $\eta(J,M)$
[Eq.~\eqref{eq:eta}], so the area is mismatched and the ratio falls to $\sim0.58$ at
$\SI{50}{\kelvin}$ for $\Jmax\ge10$, a $\sim40\%$ shortfall against the ceiling.

The physical content of Fig.~\ref{fig:sim-vs-bound}(b) is that the high-$\Jmax$
protocols lose relatively \emph{more} efficiency at high $T$ even though their absolute
orientation remains larger [Fig.~\ref{fig:sim-vs-bound}(a)]. This is exactly what the
bound anticipates through remark (ii) of Sec.~\ref{sec:bound}: at large $\Jmax$ the
ceiling approaches unity, but a single fixed-area train cannot align the dominant
$\cos\theta$ eigenvector of every $M$-sector at once, and the cost of that
incompatibility grows with the number of populated sectors.

\subsection{Operability map}
\label{sec:operability}

\begin{figure*}[t]
  \centering
  \includegraphics[width=0.92\linewidth]{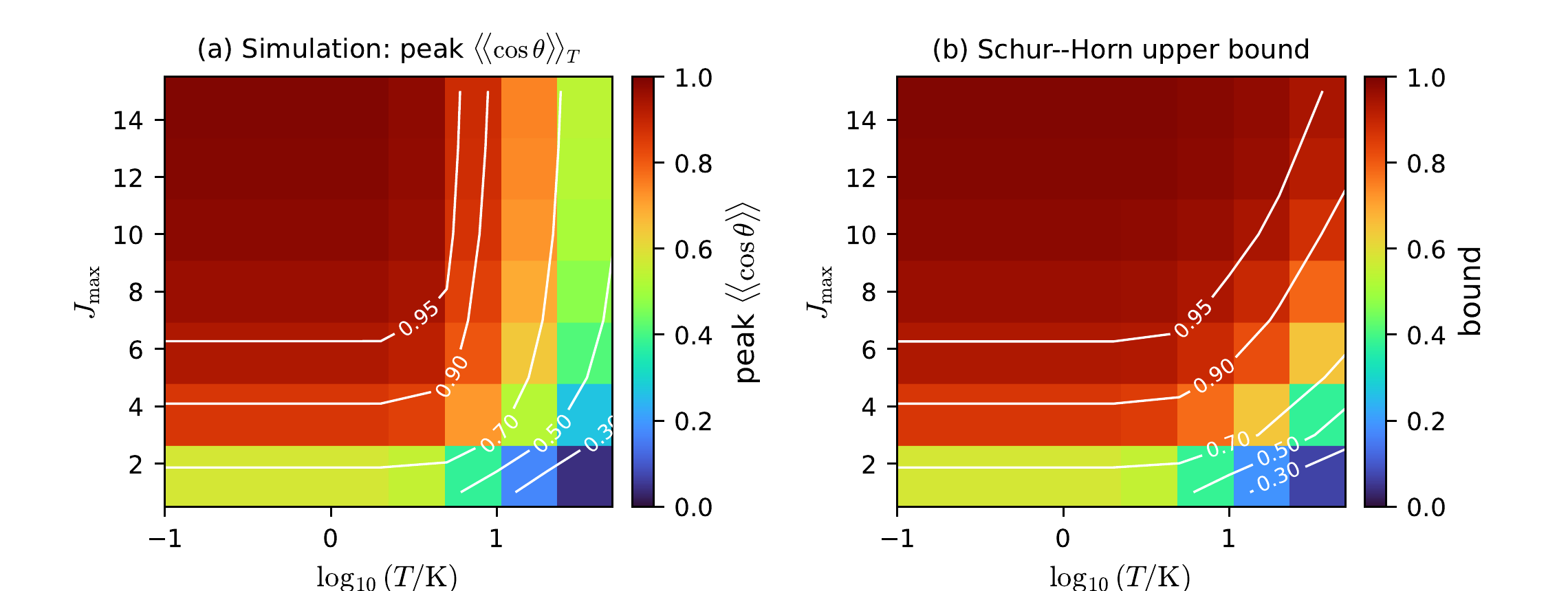}
  \caption{Operability map for LiH. (a) Simulated peak thermal orientation
  $\max_t\tavg{\cost}_T(t)$. (b) Schur--Horn upper bound of Eq.~\eqref{eq:bound}. White
  contours mark $0.30,0.50,0.70,0.90,0.95$. The two panels coincide for
  $T\lesssim\SI{5}{\kelvin}$; above this, the gap of Fig.~\ref{fig:sim-vs-bound}
  appears.}
  \label{fig:operability}
\end{figure*}

Figure~\ref{fig:operability} casts these results as a design tool: the achievable peak
orientation in the $(\Jmax,T)$ plane, alongside the bound. Reaching
$\tavg{\cost}_T\ge0.90$ requires $\Jmax\ge5$ at $\SI{5}{\kelvin}$; at $\SI{10}{\kelvin}$
it is out of reach for any $\Jmax\le15$ (the best, $\Jmax=15$, attains $0.88$), and the
shortfall only widens at higher $T$. The bound panel sets the hard limit: an optimal
control within the same $\Jmax\le15$ subspace could reach at most
$\tavg{\cost}_T\lesssim0.97$ at $\SI{20}{\kelvin}$, fixed by the partition function and
the $\cos\theta$ spectrum alone, so that even an ideal control cannot deliver strong
orientation from a $\SI{20}{\kelvin}$ ensemble without extending the controllable
subspace.

\subsection{Optimisation within the fixed pulse layout}
\label{sec:redesign}

How much of the gap to the bound is recoverable within the fixed pulse-train layout
(Gaussian widths $T_p$, central times $\tau_n$, carriers $\omega_n$)? We answer this by
optimising the thermal-peak objective directly with Nelder--Mead, seeded with the
analytic values, in two ways: varying the $N$ subpulse areas $\{\theta_n\}$ alone, and
varying the areas \emph{together with} the carrier phases $\{\phi_n\}$.
Table~\ref{tab:redesign} reports three representative cases at the demanding
temperature $T=\SI{10}{\kelvin}$. Area retuning improves the thermal peak by only
$+0.005$ to $+0.006$, closing $\approx6.5$--$7.5\%$ of the residual gap. Adding the
carrier phases yields no further gain whatsoever: the optimiser drives the phase
offsets to zero ($|\Delta\phi_n|\lesssim10^{-3}$ rad) and returns the area-only optimum
to within numerical precision.

\begin{table}[t]
\caption{Optimisation against the thermal peak $\max_t\tavg{\cost}_T$ at
$T=\SI{10}{\kelvin}$, seeded with the analytic values and holding the pulse layout
fixed, varying the $N$ subpulse areas alone (``area'') and the areas together with the
carrier phases (``area$+$phase''). ``Gap closed'' is
$(\tavg{\cost}^{\,\mathrm{opt}}_T-\tavg{\cost}^{\,\mathrm{base}}_T)/
(\tavg{\cost}^{\,\mathrm{bound}}_T-\tavg{\cost}^{\,\mathrm{base}}_T)$. The area$+$phase
optimum coincides with the area-only optimum, the optimal phase offsets vanishing
($|\Delta\phi_n|\lesssim10^{-3}$ rad), so phase freedom adds nothing within the fixed
layout.}
\label{tab:redesign}
\begin{ruledtabular}
\begin{tabular}{cccccr}
$\Jmax$ & baseline & area & area$+$phase & bound & gap closed \\
\colrule
$5$  & $0.805$ & $0.811$ & $0.811$ & $0.890$ & $7.5\%$ \\
$7$  & $0.842$ & $0.848$ & $0.848$ & $0.935$ & $6.9\%$ \\
$10$ & $0.867$ & $0.873$ & $0.873$ & $0.964$ & $6.5\%$ \\
\end{tabular}
\end{ruledtabular}
\end{table}

This near-null result localises the dominant high-temperature loss. The analytic
phases $\phi_n=\pi/2+\omega_n\tau_n$ already place the orientation maximum at the
revival, so phase offsets only misalign the amplitudes and lower the peak; within the
rigid layout (equal widths, equal spacings, resonant carriers) neither areas nor phases
offer the freedom to compensate the $M$-dependent area mismatch [Eq.~\eqref{eq:eta}]
across sectors simultaneously. The residual $\sim10$--$40\%$ gap at
$T\gtrsim\SI{10}{\kelvin}$ is therefore structural, set by the layout itself, rather
than by the choice of areas or phases. Closing it would require a richer ansatz,
relaxing the equal-width and equal-spacing constraints or admitting a full
optimal-control field with many more parameters, for example a gradient-ascent
(GRAPE) or chopped-random-basis (CRAB) optimisation~\cite{Khaneja2005,Caneva2011}
within the broader optimal-control toolbox~\cite{Glaser2015,Koch2022}, seeded by the
$M$-resolved Schur--Horn target states. The bound of Eq.~\eqref{eq:bound} furnishes the
control-independent target for any such effort.

\subsection{Experimental implications and limitations}
\label{sec:implications}

\emph{Experimental regimes.} The temperature window in which the analytic protocol is
essentially optimal coincides with several modern preparation methods. Optoelectrical
cooling reaches $\sim\SI{100}{\milli\kelvin}$~\cite{Prehn2016} and direct laser cooling
the sub-mK regime~\cite{Anderegg2017,Truppe2017}, both deep in Regime~I. Buffer-gas
sources at $\sim4$--$\SI{6}{\kelvin}$~\cite{PattersonDoyle2007} sit in Regime~II, where
the protocol retains $\ge95\%$ of its zero-temperature orientation and the bound is
essentially saturated. The challenging window is $T\gtrsim\SI{10}{\kelvin}$ (moderate
buffer-gas cells and warmer supersonic jets), where the protocol loses
$\sim10$--$40\%$; the operability map of Fig.~\ref{fig:operability} translates a target
orientation directly into the required $(\Jmax,T)$.

\emph{Reach of an improved control.} Saturating the bound requires a control whose
unitary maps the dominant Boltzmann eigenstates onto the dominant $\cos\theta$
eigenvectors in every $M$-sector at once; the fixed layout cannot
(Sec.~\ref{sec:redesign}). A layout-flexible or fully numerical design seeded by the
$M$-resolved target states is the natural route, and the bound makes its potential gain
quantitative~\cite{Karle2023,Milner2024}.

\emph{Limitations.} We have neglected (i) centrifugal distortion, which shifts the
energies by $-\Dexp J^2(J+1)^2$. For $^7$LiH the experimental distortion constant is
$\Dexp\approx\SI{8.6e-4}{\per\centi\meter}$~\cite{HuberHerzberg,NISTWebBook}, so
$\Dexp/B\approx1.1\times10^{-4}$ and the fractional shift $(\Dexp/B)J(J+1)$ stays below
$1\%$ for $J\lesssim8$ ($0.82\%$ at $J=8$, $1.03\%$ at $J=9$); at the top of the basis it
grows to $|\Delta E_{\mathrm{cd}}|\approx\SI{50}{\per\centi\meter}$ at $J=15$ (about a
fifth of the local level spacing), so for the full $\Jmax=15$ design the rigid-rotor
treatment is quantitatively reliable for the thermally dominant low-$J$ manifold
($J\lesssim5$, where the shift is $\lesssim0.4\%$ and which carries the bulk of the
population at $T\lesssim\SI{10}{\kelvin}$). Because distortion shifts only the energies,
it leaves the $\cos\theta$ spectrum, and hence the bound of Eq.~\eqref{eq:bound}, intact,
and it can be incorporated by replacing $BJ(J+1)$ with $BJ(J+1)-\Dexp J^2(J+1)^2$ in
$H_0$ at no change to the $M$-sector formalism. We have also neglected (ii) spontaneous
emission, with millisecond-scale rotational lifetimes negligible on the sub-nanosecond
protocol, and (iii) collisional dephasing, which could matter in dense supersonic jets
and which a master-equation extension would capture within the same $M$-sector
framework. Thermal averaging is thus the leading non-ideality at the temperatures of
practical interest.

\begin{figure}[t]
  \centering
  \includegraphics[width=\linewidth]{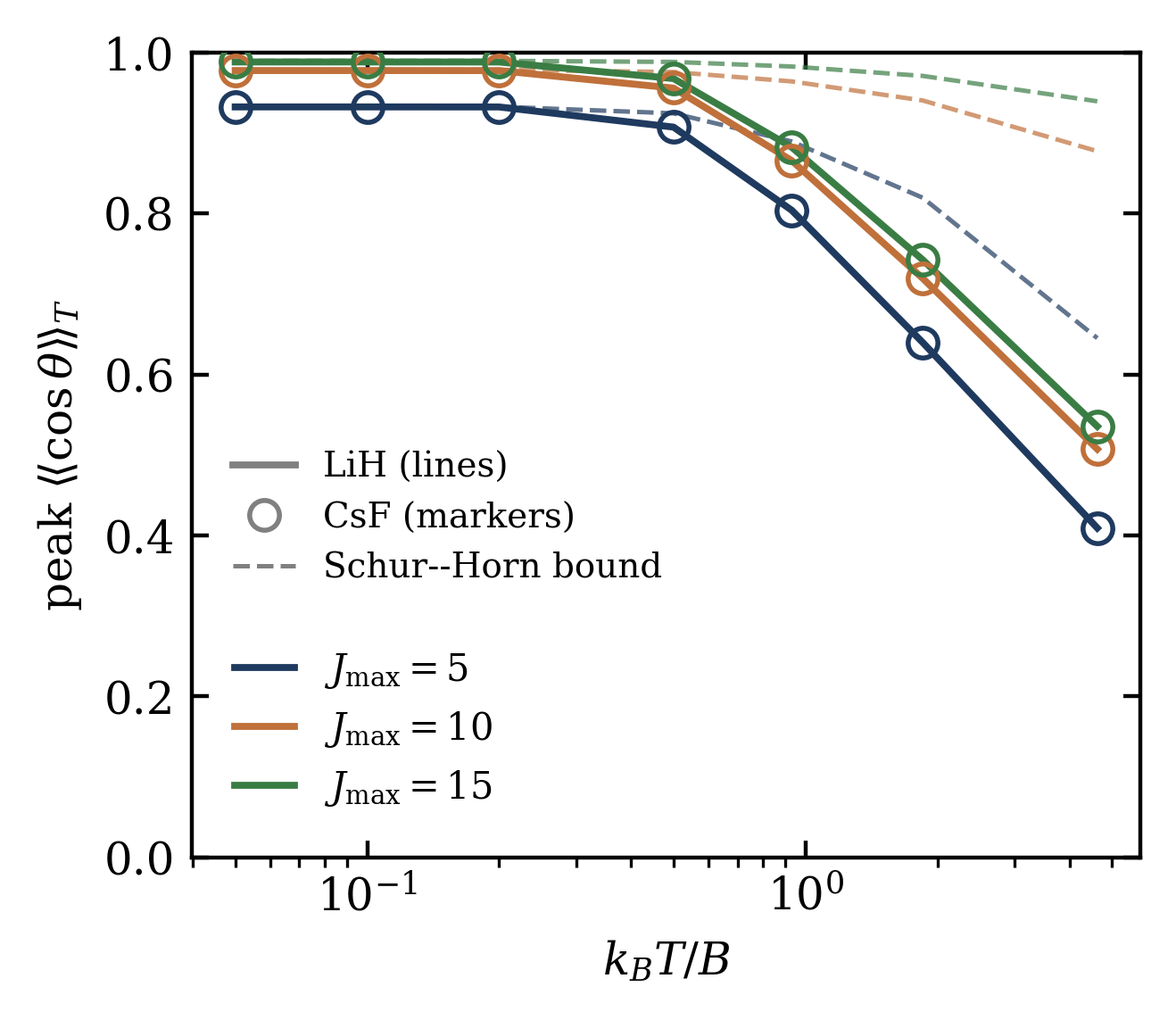}
  \caption{Molecule independence of the orientation problem. Peak thermal orientation
  $\max_t\tavg{\cost}_T$ (solid lines, LiH; open circles, CsF) and the Schur--Horn bound
  (dashed) versus the reduced temperature $\kB T/B$, for $\Jmax=5,10,15$. CsF
  ($B=\SI{0.184}{\per\centi\meter}$, $\mu_0=\SI{7.88}{D}$) has a forty-fold smaller
  rotational constant and a larger dipole than LiH, yet both protocol and bound depend
  only on $\kB T/B$: the curves coincide to $<3\times10^{-4}$ over the full range, in
  agreement with the scaling argument of the text. The bound is molecule-independent
  exactly.}
  \label{fig:universality}
\end{figure}

\emph{Generalisations.} The $M$-sector formalism and the bound apply unchanged to any
linear polar molecule in a $^1\Sigma^+$ state. In fact the molecule enters only through
the dimensionless ratio $\kB T/B$: the $\cos\theta$ spectrum is purely angular, so the
bound is molecule-independent \emph{exactly}, and because the permanent dipole cancels
in the scaled coupling ($\mu_0 F_n^{(0)}\propto\theta_n/[\eta_n T_p]$ with
$\eta_n=n/\sqrt{(2n+1)(2n-1)}$ a pure angular factor) while $B T_p$ is fixed by the
layout, the entire scaled-time dynamics, and hence the peak orientation, are
molecule-independent in the rigid-rotor model. Figure~\ref{fig:universality} confirms
this: the peak orientation and the bound for CsF ($B=\SI{0.184}{\per\centi\meter}$,
$\mu_0=\SI{7.88}{D}$), whose rotational constant is forty times smaller and whose dipole
is larger than LiH's, collapse onto the LiH curves as functions of $\kB T/B$, coinciding
to $<3\times10^{-4}$ across the full range. The only molecule-specific correction is
centrifugal distortion, whose relative size $\Dexp/B$ differs between species and which
leaves the bound untouched. Symmetric and asymmetric tops introduce additional quantum
numbers and selection rules, but the structure, a Schur--Horn bound per
conserved-quantum-number sector summed over sectors, generalises; we leave those cases
to future work.

\section{Conclusion}
\label{sec:conclusion}

We have derived an analytical Schur--Horn upper bound on the field-free orientation
attainable by any $M$-conserving unitary control acting on a thermal rotational
ensemble truncated to $J\le\Jmax$ [Eq.~\eqref{eq:bound}]. The bound is purely
kinematic, control-independent, and interpolates exactly between the zero-temperature
subspace eigenvalue and a finite-temperature ceiling set by the rotational partition
function. Using it as an absolute benchmark for a representative analytic
$N$-subpulse protocol on LiH, we found three regimes: the protocol saturates the bound
to within about $1.6\%$ (ratio $\gtrsim0.984$) for $T\le\SI{5}{\kelvin}$, loses
$\sim10\%$ of its orientation by $T=B/\kB\approx\SI{10.8}{\kelvin}$, and leaves a
$10$--$40\%$ gap above $\SI{10}{\kelvin}$. Direct optimisation of the subpulse areas and
carrier phases within the fixed layout closes only $\approx7\%$ of that gap at
$\SI{10}{\kelvin}$, with the optimal phase offsets vanishing, localising the dominant
high-temperature loss in the rigidity of the analytic layout rather than in the choice
of areas or phases, and identifying layout flexibility as the relevant next degree of
freedom. The operability map translates these limits into
experimental design rules across buffer-gas-cooled, Stark-decelerated, and
optoelectrically cooled samples, and the bound provides the rigorous target against
which any future control strategy for thermal molecular orientation can be measured.




\appendix

\section{Numerical implementation}
\label{app:numerics}

The validation of Sec.~\ref{sec:validation} uses a Strang split-operator scheme with
$\Delta t\approx\SI{38}{a.u.}$ over $240\,\Trot$ ($5.25\times10^5$ steps); the dipole
matrix $\mu_0\cos\theta$ is diagonalised once, so each step costs $O(d^2)$ for $d=21$.
For the thermal computations of Sec.~\ref{sec:thermal-response} we use the same scheme
in each $M$-sector with $\Delta t\approx\SI{220}{a.u.}$ ($1.0\times10^5$ steps over
$240\,\Trot$), propagating all initial states $\ket{J_0,M}$ within a sector in parallel
as the columns of a $d_M\times d_M$ identity, so that the per-step cost of
$\avg{\cost}$ over the full set of initial states is $O(d_M^2)$. A $T=0$
cross-validation confirms the coarser thermal grid agrees with the fine grid to
$1\times10^{-4}$ in $\max\avg{\cost}$ for all $\Jmax\le15$.

\emph{Validity of the resonant approximation.} Treating each subpulse as driving a
single $\ket{n-1,M}\to\ket{n,M}$ transition is justified because the detuning to the
nearest neighbouring transition, $2B$, greatly exceeds a subpulse bandwidth
$1/T_p=B/3\pi$: their ratio is $2B/(1/T_p)=6\pi\approx19\gg1$, so off-resonant
overlap is suppressed at the $\sim$$19$-bandwidth ($6\pi$) level.

\emph{Basis-truncation error.} The thermal basis is truncated at $J_0\le18$, three
levels above the design ceiling $\Jmax=15$. Even the population above the ceiling,
$P(J>15)$, is negligible: $\approx2\times10^{-25}$ at $\SI{50}{\kelvin}$ and
$\approx2\times10^{-63}$ at $\SI{20}{\kelvin}$. The weight at the basis edge is smaller
still (the single-level factor $e^{-18\cdot19\,B/\kB T}\approx8\times10^{-33}$ at
$\SI{50}{\kelvin}$, full tail $P(J>18)\approx2\times10^{-35}$), so the truncation
contributes no error at the $10^{-4}$ precision of the reported orientations.

\emph{Field strengths.} Table~\ref{tab:fields} lists the peak subpulse amplitudes of
Eq.~\eqref{eq:amplitude} for the $\Jmax=15$ LiH design. All lie between $6$ and
$\SI{18}{\kilo\volt\per\centi\meter}$, within reach of standard microwave/THz
pulse-train sources.

\begin{table}[t]
\caption{Peak subpulse field amplitudes $F_n^{(0)}$ [Eq.~\eqref{eq:amplitude}] for the
$\Jmax=15$ LiH design, with analytic areas $\theta_n$ and dipole matrix elements
$\mu_n$. Conversion uses $1\ \mathrm{a.u.}=\SI{5.142e9}{\volt\per\centi\meter}$.}
\label{tab:fields}
\begin{ruledtabular}
\begin{tabular}{cccc}
$n$ & $\mu_n$ (a.u.) & $\theta_n$ (rad) & $F_n^{(0)}$ (kV/cm) \\
\colrule
$1$  & $1.3359$ & $1.4540$ & $16.22$ \\
$2$  & $1.1948$ & $1.3684$ & $17.06$ \\
$3$  & $1.1733$ & $1.3085$ & $16.62$ \\
$4$  & $1.1660$ & $1.2582$ & $16.08$ \\
$5$  & $1.1627$ & $1.2127$ & $15.54$ \\
$6$  & $1.1609$ & $1.1693$ & $15.01$ \\
$7$  & $1.1599$ & $1.1263$ & $14.47$ \\
$8$  & $1.1592$ & $1.0822$ & $13.91$ \\
$9$  & $1.1587$ & $1.0355$ & $13.31$ \\
$10$ & $1.1583$ & $0.9842$ & $12.66$ \\
$11$ & $1.1581$ & $0.9259$ & $11.91$ \\
$12$ & $1.1579$ & $0.8565$ & $11.02$ \\
$13$ & $1.1578$ & $0.7693$ & $9.90$ \\
$14$ & $1.1576$ & $0.6507$ & $8.37$ \\
$15$ & $1.1575$ & $0.4681$ & $6.03$ \\
\end{tabular}
\end{ruledtabular}
\end{table}

\section{Schur--Horn bound in the \texorpdfstring{$M=0$}{M=0} sector}
\label{app:m0}

In the $M=0$ sector the bound of Eq.~\eqref{eq:bound} reduces to
\begin{equation}
  \tavg{\cost}_T^{\,\mathrm{bound},(M=0)}(\Jmax)
  =\frac{1}{Z}\sum_{J=0}^{\Jmax}
  e^{-BJ(J+1)/\kB T}\,\lambda_J^{(0)\downarrow},
  \label{eq:m0bound}
\end{equation}
with $\lambda_J^{(0)\downarrow}$ the decreasing-sorted eigenvalues of
$[\cost]_{\mathrm{sub}}^{(M=0)}$. Because this spectrum is symmetric about zero (parity
of $\cos\theta$), $\tavg{\cost}_T^{\,\mathrm{bound},(M=0)}\to0$ as $T\to\infty$, and the
leading low-$T$ correction is suppressed by $e^{-2B/\kB T}$. The general-$M$ terms in
Eq.~\eqref{eq:bound} are evaluated by direct diagonalisation of the tridiagonal
$M$-sector $\cos\theta$ matrix and summation over $M$.

For reproducibility, the $\cos\theta$ eigenvalues of the $\Jmax=5$ truncation, sector by
sector, are: $M=0$: $\pm0.93247,\ \pm0.66121,\ \pm0.23862$; $M=1$: $\pm0.83022,\ \pm0.46885,\ 0$;
$M=2$: $\pm0.69475,\ \pm0.25056$; $M=3$: $\pm0.52223,\ 0$; $M=4$: $\pm0.30151$; $M=5$: $0$. The
largest, $\lambda^{(M=0)}(\Jmax=5)=0.9324695$, is the zero-temperature orientation ceiling
for the five-subpulse design.


\end{document}